\nonstopmode
\documentclass[apl,twocolumn,reprint,nofootinbib]{revtex4-1}
\usepackage{graphicx}
\usepackage[font=small,justification=raggedright,labelfont=bf]{caption}
\usepackage[hidelinks]{hyperref}

\usepackage{amsmath, amssymb}

\begin{document}
	
\title{Nanosecond-timescale spin transfer using individual electrons in a quadruple-quantum-dot device}
\author{T.~A. Baart$^{1}$}
\author{N. Jovanovic$^{1}$}
\author{C. Reichl$^{2}$}
\author{W. Wegscheider$^{2}$}
\author{L.~M.~K. Vandersypen$^{1}$}
\affiliation{1: QuTech and Kavli Institute of Nanoscience, Delft University of Technology, P.O. Box 5046, 2600 GA Delft, The Netherlands}
\affiliation{2: Solid State Physics Laboratory, ETH Z\"{u}rich, 8093 Z\"{u}rich, Switzerland}
\date{\today}

\begin{abstract}
	The ability to coherently transport electron-spin states between different sites of gate-defined semiconductor quantum dots is an essential ingredient for a quantum-dot-based quantum computer. Previous shuttles using electrostatic gating were too slow to move an electron within the spin dephasing time across an array. Here we report a nanosecond-timescale spin transfer of individual electrons across a quadruple-quantum-dot device. Utilizing enhanced relaxation rates at a so-called `hot spot', we can upper bound the shuttle time to at most 150~ns. While actual shuttle times are likely shorter, 150~ns is already fast enough to preserve spin coherence in e.g. silicon based quantum dots. This work therefore realizes an important prerequisite for coherent spin transfer in quantum dot arrays.
\end{abstract}

\maketitle

Electrostatically defined semiconductor quantum dots have been the focus of intense research for the application of solid-state quantum computing~\cite{Hanson2007,Zwanenburg2013,Kloeffel2013}. In this architecture, quantum bits (qubits) can be defined by the spin state of an electron. Recently, several experiments have shown coherent manipulation of such spins for the purpose of spin-based quantum computation~\cite{Petta2005,Nowack2007,Medford2013,Kawakami2014,Veldhorst2014}. Enabled by advances in device technology, the number of quantum dots that can be accessed is quickly increasing from very few to several dots \cite{Thalineau2012,Takakura2014}. Large-scale quantum computing architectures require that qubits can be moved around in the course of a quantum computation~\cite{Taylor2005,Kielpinski2002}. Several approaches have been demonstrated to transfer electrons between different sites, e.g.: using surface acoustic waves~\cite{Bertrand2015a} or electrostatic gates~\cite{Baart2015}. First evidence has been shown that the spin projection is preserved during such a shuttle. It still remains to be demonstrated however that a coherent superposition is also preserved during shuttling, an essential requirement for a quantum computer. 

The approach using electrostatic gates has proven to provide high-fidelity spin transfer~\cite{Baart2015}. However, it has been challenging to create high tunnel couplings between neighbouring dots, whilst keeping sufficient coupling with nearby reservoirs to load spin states and perform the spin read-out using spin-to-charge conversion~\cite{Elzerman2004a}. In the most recent example of a shuttle~\cite{Baart2015}, the inter-dot tunnel couplings were below 1~GHz making it impossible to shuttle on the nanosecond-timescale. Given the rapid dephasing time, $T_{2}^{*}$, of $\sim$20 ns (in GaAs~\cite{Hanson2007}), such high-speed shuttles are essential to perform a coherent spin transfer. In general, short shuttle times will be beneficial.

In this Letter, we demonstrate the fast transfer of an electron-spin state inside a linear quadruple-quantum-dot device with high inter-dot tunnel couplings. To probe the spin-transfer fidelity of the shuttle we create a range of different spin states in the leftmost dot, shuttle the electron to the rightmost dot and record what happens to the spin state. Using enhanced spin-relaxation rates at a so-called `hot spot' we can upper bound the shuttle time. 

\begin{figure*}[!htb]
	\centering
	\includegraphics[width=0.90\textwidth]{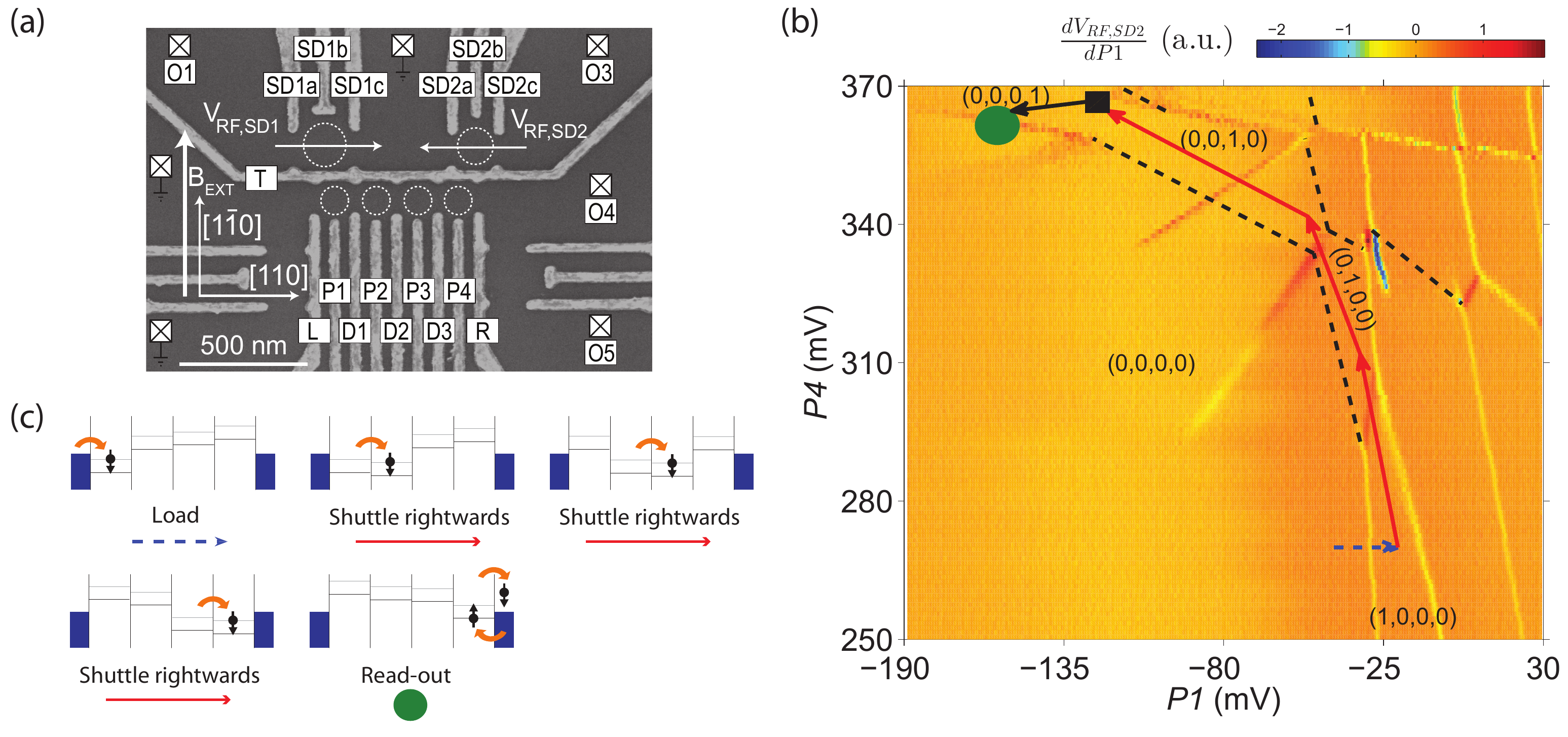}  
	\caption{(a) SEM image of a sample nominally identical to the one used for the measurements. Dotted circles indicate quantum dots, squares indicate Fermi reservoirs in the 2DEG, which are connected to ohmic contacts. The gates that are not labeled are grounded. The reflectance of SD2, $V_{RF,SD2}$, is monitored. (b) Charge stability diagram of the quadruple dot. The occupancy of each dot is denoted by $(l,m,n,p)$ corresponding to the number of electrons in dot 1 (leftmost), 2, 3 and 4 (rightmost) respectively. The fading of charge transition lines from dot 2 and 3 can be explained in a similar way as in Ref.~\citenum{Yang2014} (black dotted lines indicate their positions) and becomes less prominent for a slow scan (see Supplementary Information~\ref{sec:slow_DC_honeycomb}). The pulse sequence for loading and read-out is indicated in the charge stability diagram via arrows, see also panel b. The black rectangle corresponds to the hot spot in dot 4 where spins relax on a sub-microsecond timescale; this hot spot is only used for the measurements of Fig.~\ref{fig:figure_3}. (c) Read from left to right and top to bottom. The system is initialized by loading one electron from the left reservoir. Next, we shuttle the electron to dot 2, 3 and 4 sequentially and finally read out the spin state using spin-selective tunneling.}
	\label{fig:figure_1}
\end{figure*}

A scanning electron microscopy (SEM) image of a device nominally identical to the one used is shown in  Fig.~\ref{fig:figure_1}(a). Gate electrodes fabricated on the surface of a GaAs/AlGaAs heterostructure are biased with appropriate voltages to selectively deplete regions of the two-dimensional electron gas (2DEG) 90~nm below the surface and define the quantum dots. The main function of each gate is as follows: gates $L$ and $R$ set the tunnel coupling with the left and right reservoir, respectively. $D1-D3$ control the three inter-dot tunnel couplings and $P1-P4$ are used to set the electron number in each dot. The inter-dot tunnel couplings have each been tuned to above 2.5~GHz (see Supplementary Information~\ref{sec:inter_dot_tunnel_couplings}). We label the dots $1-4$ starting from left (1) to right (4). A nearby quantum dot on top of the qubit array, sensing dot (SD2), is created in a similar way and functions as a capacitively coupled charge sensor of the dot array. When positioned on the flank of a Coulomb peak, the conductance through the sensing dot is very sensitive to the number of charges in each of the dots in the array. Changes in conductance are measured using radiofrequency (RF) reflectometry \cite{Barthel2010}. High-frequency lines are connected via bias-tees to gates $P1$, $P3$ and $P4$. The device was cooled inside a dilution refrigerator to a base temperature of $\sim$22 mK. An in-plane magnetic field $B_{ext} = 3.5$~T was applied to split the spin-up ($\uparrow$) and spin-down ($\downarrow$) states of the electron by the Zeeman energy, thereby defining a qubit.

The spin shuttle was initialized by loading a random electron-spin from the left reservoir into dot 1, as described by the schematic diagrams of Fig.~\ref{fig:figure_1}(c) and implemented by the pulse sequence depicted by the arrows in Fig.~\ref{fig:figure_1}(b). The loading of a random electron typically results in a spin mixture of $\sim$35\% spin-$\downarrow$ and 65\% spin-$\uparrow$~\cite{Elzerman2004a}. Next, we quickly change the electrochemical potential of dot 1 and 2 in such a way that the electron will shuttle to dot 2. This is repeated for dot 2 to 3, and finally for dot 3 to 4 following the red arrows of Fig.~\ref{fig:figure_1}(b). The electrochemical potential of dot 4 is then tuned to the position of the green circle in Fig.~\ref{fig:figure_1}(b). At this position an excited spin-$\downarrow$ was allowed to tunnel to the reservoir, while a ground-state spin-$\uparrow$ would remain in the dot. The nearby sensing dot (SD2) was then used to record whether or not the electron had tunneled out, thereby revealing its spin state~\cite{Elzerman2004a}. 

\begin{figure}[!t]
	\centering
	\includegraphics[width=0.42\textwidth]{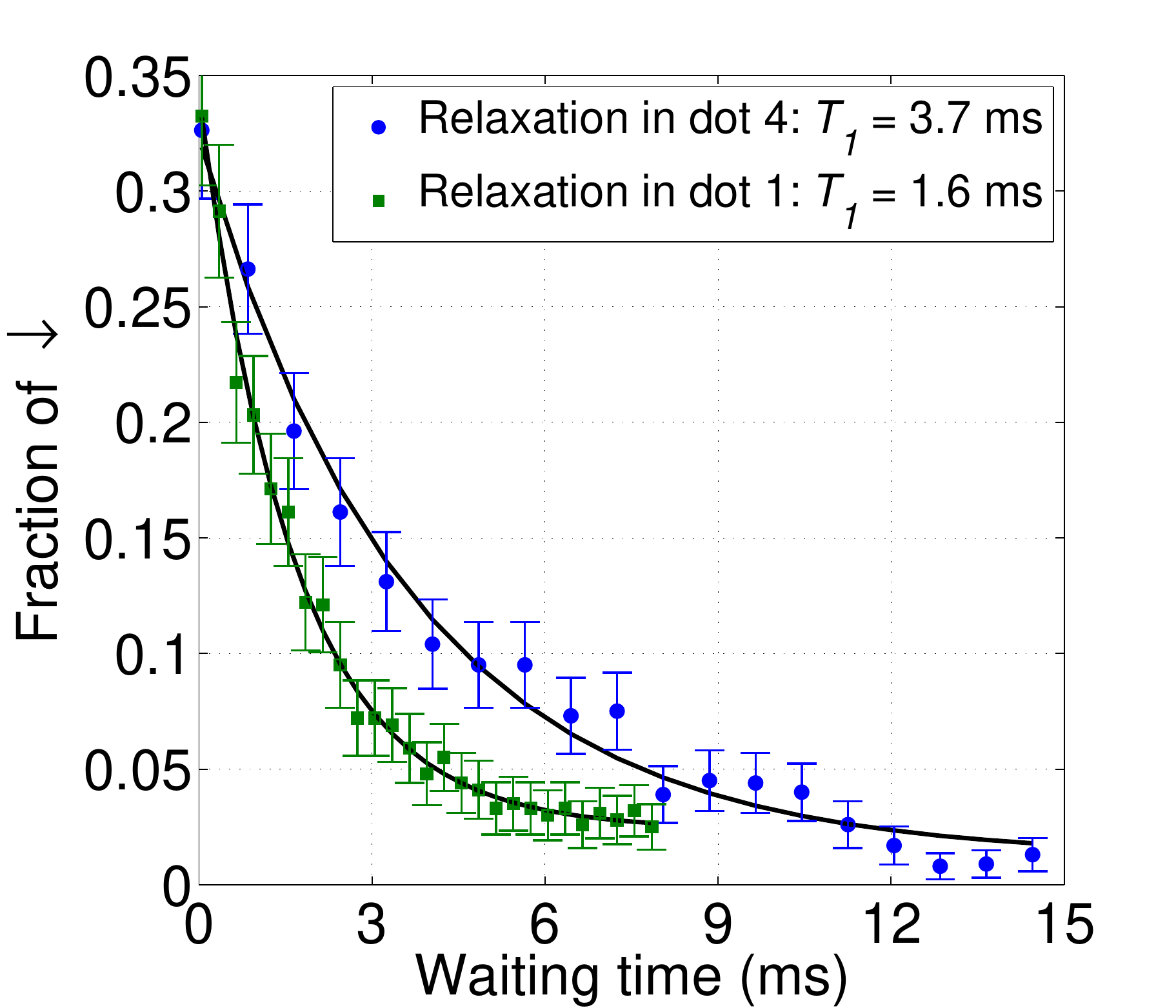}  
	\caption{A random electron-spin is loaded inside dot 1. Afterwards the spin is either allowed to relax inside dot 1 (green trace) and next shuttled to dot 4 within 3~ns where the spin is read out. Or, the electron is directly shuttled from dot 1 to 4 within 99~ns, and relaxation is induced inside dot 4 after which it is read out (blue trace). Each datapoint is an average of 999 measurements (error bars two s.d.). Solid lines indicate an exponential fit to the data of the form $P(\downarrow) = p \cdot e^{\frac{-t}{T_{1}}}+\alpha$.}
	\label{fig:figure_2}
\end{figure}

The operation of the spin shuttle was tested by performing a variety of measurements. The first two consist of introducing an extra variable waiting time inside either dot 1 or dot 4 which induces spin relaxation to the ground state spin-$\uparrow$. We will test if this is reflected in the measurement statistics. For the data represented by the blue curve in Fig.~\ref{fig:figure_2}, we first load a random electron-spin in dot 1 for 10~$\mu$s. Next we program a rectangular-shaped voltage pulse of 1~ns that induces tunneling to dot 2, then to dot 3 in 1~ns, afterwards to dot 4 in 97~ns resulting in a total shuttle time of 99~ns and add an extra waiting stage in dot 4. Finally the read-out occurs which takes 320~$\mu$s. To measure the $T_{1}$ time in dot 4 the total shuttle time is not critical as long as it is much shorter than $T_{1}$. The data shows an expected exponential decay in the measured fraction of spin-$\downarrow$ of the form $P(\downarrow) = p \cdot e^{\frac{-t}{T_{1}^{j}}}+\alpha^{j}$, where $p$ is proportional to the initial loading probability of a spin-$\downarrow$, $T_{1}^{j}$ the relaxation time in dot $j$ and $\alpha^{j}$ an offset. We observe $T_{1}^{4}=3.7~(3.1, 4.4)$~ms and $\alpha^{4} = 0.012~(0.00, 0.025)$ (values in brackets indicate 95\% confidence interval).

For the data represented by the green curve in Fig.~\ref{fig:figure_2} we perform a similar pulse sequence as before, only this time we add the extra waiting stage in dot 1 instead of dot 4. Also, the programmed shuttling time from dot 3 to dot 4 is shortened to 1~ns giving a total shuttling time of 3~ns which is close to the fastest pulse that can be applied by the used pulse generator. We observe $T_{1}^{1}=1.6~(1.5, 1.8)$~ms and $\alpha^{1} = 0.024~(0.019, 0.030)$. The reported values for $T_{1}$ are in correspondence with earlier measurements~\cite{Hanson2007}. 

An important ingredient of a spin-shuttle is preservation of the spin state during a shuttle. This state could be influenced whilst shuttling due to a variety of mechanisms: (1) charge exchange with the reservoirs, (2) spin-orbit (SO) interaction and (3) hyperfine interaction with the nuclear spins of the quantum-dot host material. A detailed discussion is given in Ref.~\citenum{Baart2015}. To determine if spin flips occur, we can compare the value of $\alpha^{1}$ and $\alpha^{4}$. For the $T_{1}$ measurement in dot 4, $\alpha^{4}$ corresponds to `1 minus the spin-$\uparrow$ read-out fidelity', assuming perfect spin-$\uparrow$ initialization by thermalization~\cite{Baart2015}. $\alpha^{1}$ describes the probability to measure a spin-$\downarrow$ in dot 4, after having created a spin-$\uparrow$ in dot 1 by waiting infinitely long. The read-out fidelity does not depend on in which dot the $T_{1}$ process is induced, or on the shuttling time from dot 1 to 4. As a consequence, the value of $\alpha^{1}$ can be used to determine if spin flips have occurred as a spin-$\uparrow$ from dot 1 is shuttled to dot 4. If $\alpha^{1}$ is larger than $\alpha^{4}$, this would indicate spin flips. Since the confidence intervals for $\alpha^{1}$ and $\alpha^{4}$ overlap, we conclude that there is no evidence for spin flips during shuttling.

The measurements so far strongly indicate that we have good control over where the electron spin resides (different $T_{1}$'s), and that no spin-flips are induced even when shuttling at high speed throughout the array (similar $\alpha$'s). However, due to the relatively long read-out time of 320~$\mu$s it was still possible that even though we \textit{programmed} a pulse sequence that should correspond to a shuttle time of 3~ns from dot 1 to 4, the electron actually remained for a longer time in one of the dot(s) $1-3$ and only shuttled to dot 4 somewhere during the read-out stage. During such an event, the electron lagging behind would temporarily be in a dot whose electrochemical potential is \textit{above} the Fermi-level of the reservoirs and has sufficient energy to potentially leave the dot array. If it leaves, a new random electron will enter the array which would be detrimental for the shuttle-functionality. Alternatively, the electron stays within the array and continues the shuttle to end up in dot 4. 

\begin{figure}[!tb]
	\centering
	\includegraphics[width=0.35\textwidth]{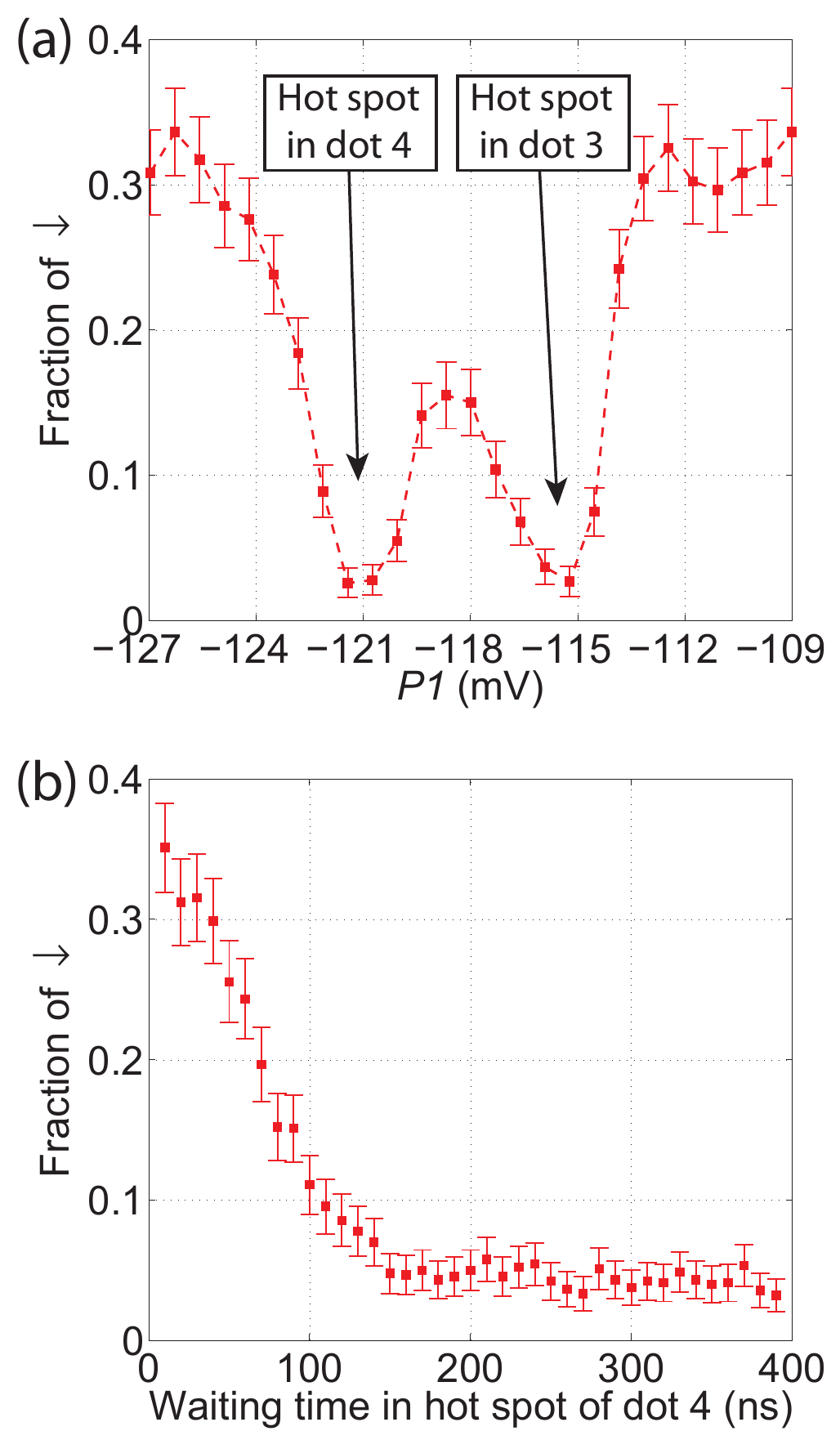}  
	\caption{(a) A random electron-spin is loaded inside dot 1. It is then shuttled to dot 2 for 1~ns, and next to dot 3 for 1~ns. Afterwards we vary $P_{1}$ across the inter-dot transition of dot 3 and 4 and wait there for 500~ns to identify the location of the hot spots. Each datapoint is an average of 999 measurements (error bars two s.d.). Dashed line is a guide to the eye. (b) Same pulse sequence as (a), only this time we pulse to the hot spot in dot 4 determined from (a) ($P_{1} \approx -121$~mV), and vary the waiting time inside the hot spot. Each datapoint is an average of 900 measurements (error bars two s.d.). The spin state quickly relaxes due to the short $T_{1}$-time of the hot spot.}
	\label{fig:figure_3}
\end{figure}

To gain more insight in when the electron actually arrives in dot 4, we added an extra stage to the pulse sequence that corresponds to a so-called `hot spot' in dot 4. At the location of this hot spot, spin-orbit and hyperfine interactions rapidly mix the spin excited state `$\downarrow(0,0,0,1)$' with the orbital excited state `$\uparrow(0,0,1,0)$'. Spin-conserving orbital relaxation quickly transfers the state `$\uparrow(0,0,1,0)$' to `$\uparrow(0,0,0,1)$'. As a result, the `$\downarrow(0,0,0,1)$' state will relax on a sub-microsecond timescale to the ground state `$\uparrow(0,0,0,1)$'~\cite{Srinivasa2013}. We will now use a similar pulse sequence as before, only with the inclusion of the hot spot inside dot 4. If the electron spin indeed follows the prescribed pulse sequence, it will hit the hot spot and relax. If the electron however resides for some more time in dot $1-3$, it will afterwards miss the hot spot and will still show a significant spin-$\downarrow$ fraction. To identify the location of the hot spots we again load a random electron-spin in dot 1, pulse to dot 2 in 1~ns, then to dot 3 in 1~ns and next to a varying location along the inter-dot transition of dot 3 and 4 for 500~ns; the result is shown in Fig.~\ref{fig:figure_3}(a). This clearly shows two prominent locations where the spin has completely relaxed. The dip at $P_{1} \approx -121$~mV corresponds to the hot spot in dot 4, and is depicted by a black rectangle in Fig.~\ref{fig:figure_1}(b). The other dip ($P_{1} \approx -115$~mV) corresponds to the hot spot in dot 3 and is not used in this experiment. 

To get an upper bound for the shuttling speed we apply a sequence where we again hop from dot 1 to 2, and dot 2 to 3 in 1~ns each, and then pulse to the hot spot in dot 4 for a varying amount of time; the result is shown in Fig.~\ref{fig:figure_3}(b). This shows that after waiting 150~ns, the whole spin state has relaxed resulting in an upper bound of shuttling of $\sim$150~ns. For quantifying this upper bound we are now limited by the relaxation time of the hot spot in dot 4. We have verified that by using just dot 4 (i.e. load in dot 4, move to hot spot, and read-out from dot 4) that $\sim$150~ns is the fastest relaxation time of this hot spot. 

The upper bound of 150~ns is not yet enough to guarantee that a coherent spin transfer can be performed inside a GaAs device with a $T_{2}^{*} \approx 20$~ns. It is however promising that the spin shuttle seems to function without loss of spin-information for a pulse time as short as 3~ns such as shown in Fig.~\ref{fig:figure_2}, indicating that a coherent transfer could already be feasible in this system. This is in agreement with the observation that all inter-dot tunnel couplings exceed 2.5~GHz. In practice, the tunneling may be happening on the timescale of the rise time of the pulse, such that the tunnel events could be adiabatic with respect to the inter-dot anticrossings. In each case, 150~ns is fast enough in different host materials such as Si or Si/SiGe where this shuttling technique could in principle also be applied and the dephasing time has been measured to be much longer $>$120~$\mu$s~\cite{Veldhorst2014}. It still has to be shown that such structures can reach high tunnel couplings, although first steps have recently been made in a triple-quantum-dot device~\cite{Eng2015}. 

In summary, we have demonstrated a spin shuttle inside a quadruple-quantum-dot device where an electron-spin is shuttled within at most 150~ns across the four dots. This work forms the next step in performing a spin shuttle using electrostatic gates that demonstrates preservation of a quantum superposition, an essential ingredient for powerful quantum computing architectures.  

\begin{acknowledgments}
The authors acknowledge useful discussions with the members of the Delft spin qubit team, and experimental assistance from M.~Ammerlaan, J.~Haanstra, R.~Roeleveld, R.~Schouten, M.~Tiggelman and R.~Vermeulen. This work is supported by the Netherlands Organization of Scientific Research (NWO) Graduate Program, the Intelligence Advanced Research Projects Activity (IARPA) Multi-Qubit Coherent Operations (MQCO) Program and the Swiss National Science Foundation.		
\end{acknowledgments}

\clearpage

\begin{center}
\textbf{Supplementary Material}
\end{center}

\section{Methods and materials}
The experiment was performed on a $\mathrm{GaAs/Al_{0.307}Ga_{0.693}As}$ heterostructure grown by molecular-beam epitaxy, with a 90-nm-deep 2DEG with an electron density of $\mathrm{2.2 \cdot 10^{11}\ cm^{-2}}$ and mobility of $\mathrm{3.4 \cdot 10^{6}\ cm^{2} V^{-1} s^{-1}}$ (measured at 1.3 K). The metallic (Ti-Au) surface gates were fabricated using electron-beam lithography. The device was cooled inside an Oxford Triton 400 dilution refrigerator to a  base temperature of $\sim$22~mK. To reduce charge noise the sample was cooled while applying a positive voltage on all gates (ranging between 100 and 400~mV) \cite{Long2006}. Gates $P1$, $P3$ and $P4$ were connected to homebuilt bias-tees ($RC$=470 ms), enabling application of d.c.~voltage bias as well as high-frequency voltage excitation to these gates. Frequency multiplexing combined with RF reflectometry of the SDs was performed using LC circuits matching a carrier wave of frequency 81.0~MHz for SD2. The inductors are formed by microfabricated NbTiN superconducting spiral inductors with an inductance of 4.6~$\mu$H for SD2. The power of the carrier wave arriving at the sample was estimated to be -103 dBm. The carrier signal was only unblanked during readout. The reflected signal was amplified using a cryogenic Weinreb CITLF2 amplifier and subsequently demodulated using homebuilt electronics. Real-time data acquisition was performed using a field-programmable gate array (FPGA DE0-Nano Terasic) programmed to detect tunnel events using a Schmitt trigger. Voltage pulses to the gates were applied using a Tektronix AWG5014. Microwaves were generated using a Rohde \& Schwarz SMR40 generator connected to $P_{3}$ via a homemade bias-tee at room temperature.

\section{Charge stability diagram measured on a slow timescale}
\label{sec:slow_DC_honeycomb}
\begin{figure}[!htb]
	\centering
	\includegraphics[width=0.5\textwidth]{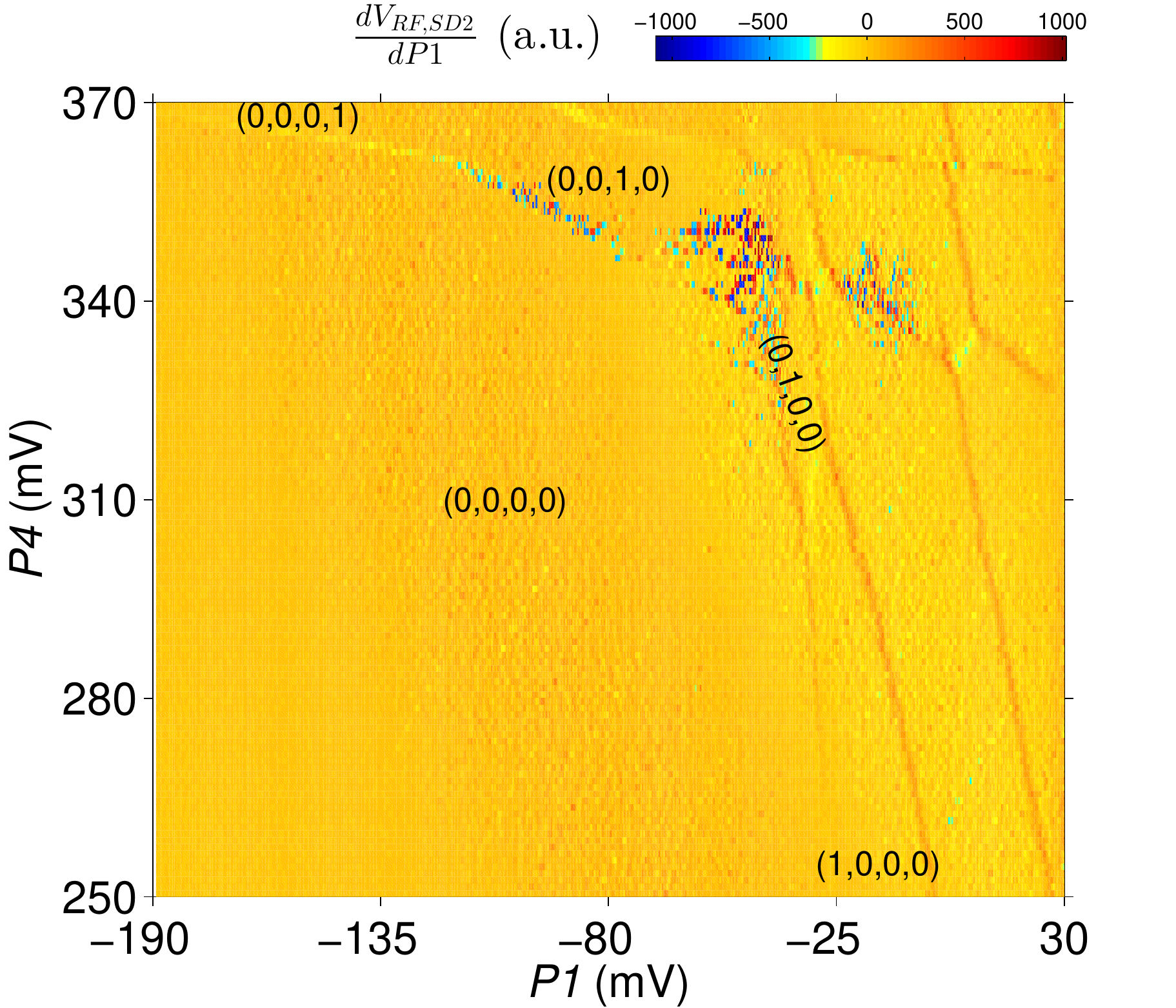}  
	\caption{Charge stability diagram as in Fig.~\ref{fig:figure_1}(b) of the main text, only this time measured in a `slow' mode where both gates are stepped using a DAC. The additional lines inside (0,0,0,0) of Fig.~\ref{fig:figure_1}(b) of the main text are not visible anymore. This measurement took $\sim$8~hours.}
	\label{figS:DC_honeycomb}
\end{figure}

The charge stability diagram shown in Fig.~\ref{fig:figure_1}(b) of the main text has been taken in a so-called `fast-honeycomb' mode~\cite{Baart2015}. Using the bias-tees connected to $P_{1}$, $P_{3}$ and $P_{4}$ it is possible to step one of them `slowly' using a DAC and apply a triangular ramp on the other using the AWG. This significantly speeds up the measurements compared to stepping both gates using DACs. In Fig.~\ref{fig:figure_1}(b) of the main text we plot the reverse sweep, i.e. the voltage on the $x$-axis is swept from positive to negative (with a rate of 220 mV/ 4.4 ms). The fading of the charging lines of dot 2 and 3 can then be explained from the indirect coupling with a reservoir~\cite{Yang2014}. To verify that this is correct, we have also measured Fig.~\ref{fig:figure_1}(b) in a slow mode where we step both gates using a DAC, the result is shown in Fig.~\ref{figS:DC_honeycomb}. 

\section{Measurements of the inter-dot tunnel couplings}
\label{sec:inter_dot_tunnel_couplings}
\begin{figure}[!htb]
	\centering
	\includegraphics[width=0.4\textwidth]{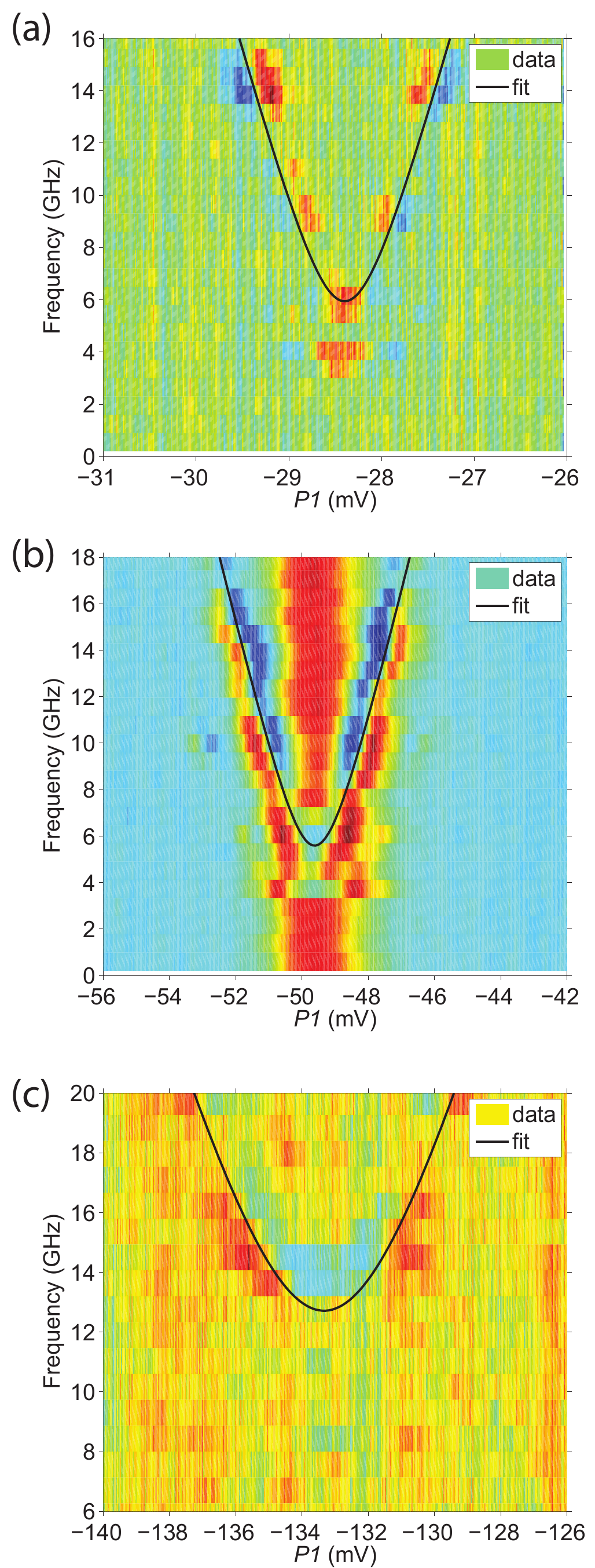}  
	\caption{(a-c) Measurements of the tunnel coupling at zero detuning between neighbouring dots. (a) PAT between (1,0,0,0)-(0,1,0,0) resulting in $t_{1,2} \approx 3.0$~GHz. (b) PAT between (0,1,0,0)-(0,0,1,0) resulting in $t_{2,3} \approx 2.8$~GHz. (c) PAT between (0,0,1,0)-(0,0,0,1) resulting in $t_{3,4} \approx 6.4$~GHz.}
	\label{figS:PAT}
\end{figure}

The tunnel coupling at zero detuning between neighbouring dots was measured using photon-assisted tunneling (PAT)~\cite{Oosterkamp1998a}, see Fig.~\ref{figS:PAT}. The data is fitted to $\sqrt{((P_{1}-P_{\mathrm{1,offset}}) \cdot \alpha_{P_{1}})^2+4t^{2}}$ where $\alpha_{P_{1}}$ is the lever arm that is different for each inter-dot transition (not used in this experiment). 

\end{document}